\documentclass[12pt]{article}
\usepackage[top = 2 cm, bottom = 3 cm, left = 3 cm, right = 2 cm]{geometry}
\usepackage{authblk, cite, color, amssymb, amsmath, epsfig}
\usepackage[colorlinks = true, linkcolor = blue, citecolor = red]{hyperref}
\usepackage[dvipsnames]{xcolor}

\begin{document}

\title{\huge Diffractions from the brane and GW150914}

\author{\bf Merab Gogberashvili}
\affil{\small Javakhishvili Tbilisi State University, 3 Chavchavadze Avenue, Tbilisi 0179, Georgia \authorcr
Andronikashvili Institute of Physics, 6 Tamarashvili Street, Tbilisi 0177, Georgia}

\author{\bf Pavle Midodashvili}
\affil{\small Ilia State University, 3/5 Cholokashvili Avenue, Tbilisi 0162, Georgia}

\maketitle

\begin{abstract}
In the braneworld scenario the zero mode gravitons are trapped on a brane due to non-linear warping effect, so that gravitational waves can reflect from the brane walls. If the reflected waves form an interference pattern on the brane then it can be detected on existing detectors due to spatial variations of intensity in the pattern. As an example we interpret the LIGO event GW150914 as a manifestation of such interference pattern produced by the burst gravitational waves, emitted by a powerful source inside or outside the brane and reflected from the brane walls.
\vskip 3mm
PACS numbers: 04.30.Nk; 98.70.Rz; 11.25.Uv
\vskip 1mm

Keywords: Gravitational waves; brane; diffraction
\end{abstract}
\vskip 5mm


Recently LIGO collaboration announced the event GW150914, which has been interpreted as the first direct observation of gravitational waves (GW) from inspiral and merger of a pair of black holes at a luminosity distance of about $0.4$ Gpc \cite{ligoGW}. For this very peculiar source no other instrument has found an associated electromagnetic signal \cite{int, des, fermiLAT, swift}, as well as no success has been achieved in high energy neutrino follow-up detection \cite{icecube}. Only the FERMI Gamma Ray Burst Monitor (by the way, having ill-constrained localization with respect to GW150914) revealed the presence of a weak hard X-ray transient $0.4$ seconds after the GW event \cite{fermiTransient}. However, the association of the transient with the GW event is not clear due to the large uncertainty in the localization of the merger and the lack of a detailed analysis of the parameters (such as the strength of the relevant magnetic fields) at the source. So room is still left for alternative interpretations of a source of the GW150914. For instance, there has been proposed hypotheses that the merging objects were gravastars \cite{gravistar}, or dark matter black holes \cite{PBH}, instead of ordinary black holes.

In this note we exploit more radical possibility that GW150914 signal was caused by a short-duration ("bursting") source within the braneworld scenario \cite{brane}, or by the local kick on the brane from the bulk GWs in the standing wave braneworld model \cite{SW-brane}. We suppose that high frequency burst GW could be reflected from the brane walls to produce temporal interference pattern in some region along the brane which almost simultaneously were passed by the two LIGO detectors located on the same continent on the Earth.

It is known that GWs are space-time fluctuations and usually they don't reflect from any matter surface. In the braneworld scenario zero mode gravitons are trapped on a brane due to non-linear warping effect, and so the ultra short GWs can be reflected from the brane walls. The amplitude of high frequency GWs is unobservable for the existing GW detectors. But if GWs form an interference pattern, as due to reflections on the brane in our case, they can be detected by exploring the spatial variations of intensity in the pattern (the second order effect in spacetime fluctuations), just like in the case of electromagnetic waves. Electromagnetic and gravitational waves are very different in many aspects. For example, GW detectors are sensitive to the amplitude of the signal, while the electro-magnetic waves are mostly detected by their intensity. However, as both are wave phenomena, we assume that electro-magnetic wave diffraction rules can be used also for GWs \cite{Torn}. Although there is no true local measure of the energy in the gravitational field, we think that in the weak field limit it's possible to experimentally measure the intensity of GWs,
\begin{equation} \label{I=A2}
I \sim A^2 ~,
\end{equation}
where $A \ll 1$ is the amplitude of the plane GWs,
\begin{equation} \label{h}
h_{\mu\nu} = A ~C_{\mu\nu} \sin (k_\sigma x^\sigma) ~.
\end{equation}
Here $k^\sigma$ is the wave vector and $C_{\mu\nu}$ is the orthogonal to $k^\sigma$ unit polarization tensor (with two independent components). Using the expressions for averages of harmonic functions:
\begin{equation} \label{sin}
\langle\sin^m (\omega x)\rangle = \frac{\omega }{2\pi} \int\limits_0^{2\pi/\omega} \sin^m (\omega x) dx = \left \{
\begin{array} {lr}
0 & (m = 2n + 1)\\
2^{-2n}\left( 2n \right)! \left( n! \right)^{-2} & (m = 2n)
\end{array}\right. ~,
\end{equation}
it's easy to see that in the first order of (\ref{h}) GW detectors are not sensitive to the waves with high frequency,
\begin{equation}
\langle h_{\mu\nu} \rangle \to 0 ~.
\end{equation}
In order to keep track of the energy carried by the gravitational waves, the metric must be expanded to at least second order in $h_{\mu\nu}$. Due to (\ref{sin}), in the second order there is the small constant distortion of the flat background metric $\eta_{\mu\nu}$,
\begin{equation}
\langle h^2_{\mu\nu}\rangle \sim A^2 ~.
\end{equation}
Usually constant shift of a metric is equivalent to the redefinition of coordinates and is unobservable. However, crossing the region with interference pattern of GWs one can measure differences in intensity (\ref{I=A2}).

In order to speak about the flat GWs its wavelength $\lambda$ should be much smaller than the radius of curvature of the background spacetime. In the braneworld scenario, to consider reflections of GWs from the brane walls, it should be also assumed that
\begin{equation}
\lambda \ll \varepsilon \sim \frac {1}{\sqrt \Lambda} ~,
\end{equation}
where $\varepsilon $ is the brane width and $\Lambda$ is the bulk cosmological constant. In general, high energy zero mode gravitons can penetrate into the bulk, scatter on the both brane 'walls' and produce an interference pattern on the brane. In the vicinity of brane walls the real part of the refraction index obeys the condition:
\begin{equation}
|n| = \frac cv \sim 1 + \epsilon > 1~,
\end{equation}
where $v$ is the effective velocity of waves and $\epsilon$ corresponds to the penetration depth towards the extra spatial dimensions. For the distances of the order of $\varepsilon$ the parameter $\epsilon$ will be significantly different from zero, since, because of warping, the effective speed of zero mode fields in the bulk becomes zero,
\begin{equation}
v = \frac {c}{|n|} \sim c( 1 - \epsilon) \to 0 ~.
\end{equation}

The situation with penetration of energetic wave into a reflecting surface reminds the case of X-ray diffraction in 4D, where one can observe asymmetric interference picture \cite{X-ray}. Using analogy with X-rays, for the diffraction angle of GWs reflected from the brane walls we use the non-linear law:
\begin{equation}\label{phi^2}
\phi^2 = \theta^2 + \delta ~.
\end{equation}
Here $\delta$ is the asymmetric diffraction parameter, and the ordinary diffraction angle $\theta$, which obeys the rules of geometrical optics, is connected with the distance $L$ to the source as,
\begin{equation} \label{theta}
\theta \sim \frac {x}{L}~,
\end{equation}
where $x$ is the coordinate along the interference pattern. For simplicity we assume that the distorted angle $\phi$ in (\ref{phi^2}) also takes into account misorientation of diffraction planes with respect to the brane's surface normal, and it corresponds to the interference fringes of reflected waves along the brane.

For the distribution of intensity in the interference pattern produced by a single-slit (brane in our case) diffraction of GWs (\ref{h}) we use the familiar formula from optics,
\begin{equation} \label{I}
I(x) = A^2 \left(\frac {\sin \beta}{\beta}\right)^2 ~.
\end{equation}
Here
\begin{equation}
\beta (x) = \frac {\pi \varepsilon}{\lambda} \sin \phi~,
\end{equation}
where according to (\ref{phi^2}) and (\ref{theta})
\begin{equation}
\phi = \sqrt {\frac{x^2}{L^2} + \delta}~.
\end{equation}
It should be noted that the zero reflection angle ($\theta \sim x = 0$) doesn't correspond to the maximum in the intensity distribution (\ref{I}), since
\begin{equation}\label{beta-theta-zero}
  \beta (0) = \frac{\pi \varepsilon}{\lambda} \sin \sqrt \delta~.
\end{equation}
Let's introduce the function
\begin{equation}\label{DeltaBeta-theta}
  \Delta _\beta (\theta ) = \beta (\theta ) - \beta (0) = \frac{\pi\varepsilon}{\lambda }\left( \sin \sqrt {\theta ^2 + \delta } - \sin \sqrt \delta  \right) ~,
\end{equation}
which defines the change of $\beta$ in terms of the the angle (\ref{theta}) along the interference pattern. For the angles $\theta_N$ of the minima of intensity distribution in the interference pattern we have
\begin{equation}\label{theta-n}
  \Delta _\beta \left( \theta_N \right) =\pi N ~,
\end{equation}
where $N$ is a positive integer. The equation (\ref{theta-n}) can be solved with sufficient accuracy if we expand $\Delta _\beta (\theta )$ in the form:
\begin{equation}\label{SeriesDeltaBeta}
\Delta _\beta (\theta ) = \frac{\pi\varepsilon}{\lambda} \left[\frac{\cos \sqrt\delta}{2\sqrt\delta }~\theta^2 - \frac{1}{8\delta} \left(\sin \sqrt\delta + \frac{\cos \sqrt \delta }{\sqrt \delta } \right)\theta ^4 + ...\right] ~.
\end{equation}
and use only the first term, proportional to $\theta^2$. So from (\ref{theta-n}) we'll have:
\begin{equation}\label{expr-theta-n}
\frac{\pi \varepsilon \cos \sqrt \delta }{2\sqrt \delta \lambda  }\theta ^2_N = \pi N ~.
\end{equation}
For the ratio of the wavelength of GWs and brane width we suppose,
\begin{equation} \label{theta-1}
\frac {\lambda}{\varepsilon} \sim \frac {M_H}{M_{Pl}} \approx 10^{-15}~,
\end{equation}
where $M_H$ and $M_{Pl}$ are the Higgs and Planck scales, respectively. Using estimations from the X-rays physics, where $\delta \sim \theta \cdot 10^{-3}$ \cite{X-ray}, for the asymmetric diffraction parameter we take:
\begin{equation} \label{delta}
\delta = 10^{-18} ~.
\end{equation}
Then using (\ref{theta}) and (\ref{expr-theta-n}) for the coordinates of the minima of intensity in the interference pattern we find:
\begin{equation}\label{solution-theta-n}
{x_N} \approx \sqrt {\frac{{2\sqrt \delta \lambda N}}{\varepsilon }} L~.
\end{equation}

We want to explain the eight cycles of the GW150914 signal lasted over $\tau = 0.2$ seconds as a result of the Earth's passing through the interference pattern (\ref{solution-theta-n}) with the velocity $v_\oplus$ relative to the pattern. For the first eight fringes from (\ref{solution-theta-n}) we find,
\begin{equation}\label{v}
v_ \oplus \approx \frac{x_8}{\tau } \approx 4 \sqrt {\frac{\sqrt \delta \lambda }{\varepsilon }} \frac L\tau ~.
\end{equation}
If we suppose that the value of the relative velocity is of the order of
\begin{equation}
v_\oplus \sim 10^3 ~km/sec~,
\end{equation}
from (\ref{v}), using the values (\ref{theta-1}) and (\ref{delta}), we calculate the distance to the source of GWs,
\begin{equation} \label{L}
L \approx \frac 14\sqrt {\frac{\varepsilon }{\sqrt \delta \lambda }} ~v_\oplus \tau \approx 5 \times 10^{13} ~km \sim 1~pc ~.
\end{equation}

\begin{figure}
  \centerline{
  \psfig{figure=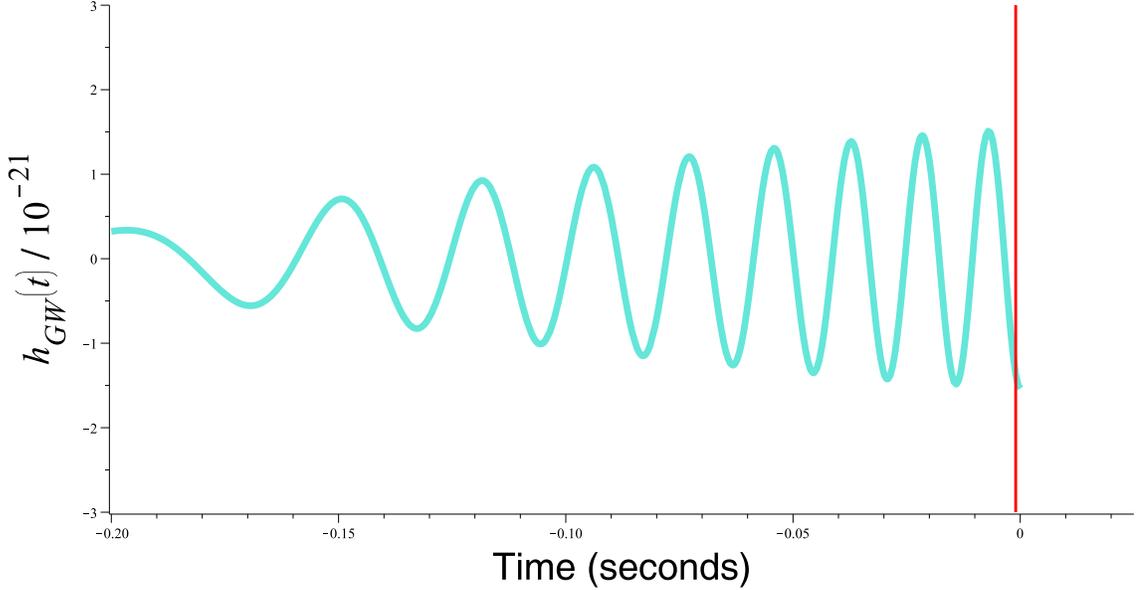,angle=0,height=8cm}}
  \caption{Impact of the GWs interference pattern on LIGO}
\label{1}
\end{figure}

Now let us estimate the relative shift of moving with the velocity of the Earth, $v_\oplus$, two LIGO mirrors at
\begin{equation}\label{LIGO-Earth}
x_1 = tv_\oplus~, ~~~~~ x_2 = tv_\oplus - l
\end{equation}
($l \approx 5 ~km$ is the distance between the mirrors), by evaluating the function:
\begin{equation}\label{LIGO-measurement}
h_{GW}(t) = \frac{\Delta l}{l} = I(x_1) - I(x_2)~,
\end{equation}
where the intensities of GWs, $I(x)$, are expressed using (\ref{I}) and (\ref{LIGO-Earth}). If for the amplitude of the initial GWs (\ref{h}) we take,
\begin{equation}
A = 10^{-4} \ll 1 ~,
\end{equation}
relative oscillations of the LIGO mirrors described by (\ref{LIGO-measurement}), for first eight fringes, is displayed in Figure \ref{1}. In our case of asymmetric diffraction, the fringe width decreases, and, correspondingly, the frequency of the signal on LIGO detector increases. For chosen values of parameters the fringe width for the first eight fringes decreases from approximately $50 ~km$ to $10 ~km$, i.e. becomes compatible with LIGO detector's size and signal becomes undistinguishable from the background noise (at the right from the red line on Figure \ref{1}).

To conclude, in this paper we have shown that the passage of the Earth through the interference pattern of burst GWs reflected from the brane can imitate the recent LIGO signal, which previously was interpreted as GWs from inspiral binary black holes.


\end{document}